\documentclass[conference]{IEEEtran}
\IEEEoverridecommandlockouts
\usepackage{cite}
\usepackage{amsmath,amssymb,amsfonts}
\usepackage{algorithmic}
\usepackage{graphicx}
 \usepackage{multirow}
 \usepackage{tablefootnote}
\usepackage{textcomp}
\usepackage{url}
\usepackage{tabularx}
\usepackage{authblk} %
\usepackage{siunitx}
\usepackage{threeparttable} 
\usepackage{fancyvrb} 
\usepackage{xcolor}
\usepackage{tikz}
\def\BibTeX{{\rm B\kern-.05em{\sc i\kern-.025em b}\kern-.08em
    T\kern-.1667em\lower.7ex\hbox{E}\kern-.125emX}}
\begin{document}
\newcommand{\Cross}{\mathbin{\tikz [x=1.4ex,y=1.4ex,line width=.2ex] \draw (0,0) -- (1,1) (0,1) -- (1,0);}}
\title{Starlink in Northern Europe: A New Look at Stationary and In-motion Performance}

\author{Muhammad Asad Ullah\textsuperscript{1}, Antti Heikkinen\textsuperscript{1}, Mikko Uitto\textsuperscript{1}, Marko Höyhtyä\textsuperscript{2}, Antti Anttonen\textsuperscript{1},\\ Konstantin Mikhaylov\textsuperscript{3},  Timo Lind\textsuperscript{1}}
\affil{ \textsuperscript{1}VTT Technical Research Centre of Finland Ltd., Espoo, Finland\\
\textsuperscript{2} Finnish National Defence University, Helsinki, Finland\\
\textsuperscript{3} Centre for Wireless Communications,  University of Oulu, Oulu, Finland\thanks{This work has been submitted to the 2025 IEEE Vehicular Networking Conference (IEEE VNC). Copyright to IEEE may be transferred without notice.}}

\maketitle
\begin{abstract}
Starlink has introduced the Flat High Performance~(FHP) terminal, specifically designed to support the vehicles and the vessels in motion as well as the high-demand stationary users. The research on FHP terminal throughput analysis remains limited, only a few existing studies evaluate FHP, focusing on the limited parameters and scenarios. This paper evaluates the FHP terminal’s performance in Finland, Northern Europe. We examine round-trip time (RTT), uplink, and downlink throughput for both stationary and in-motion use. We measure network efficiency across six geographically diverse servers and get insights of network routing strategies. Our results show that Starlink provides high-speed, low-RTT connectivity, however, the throughput experiences fluctuations with slight degradation when in motion. Additionally, we compare Starlink and terrestrial network RTT and possible routing paths.
\end{abstract}
\begin{IEEEkeywords}
ESIM, latency, mobility, satellite, throughput
\end{IEEEkeywords}

\section{Introduction}
The low Earth orbit (LEO) satellite communication sector has experienced remarkable growth, driven by the increasing demand for global wireless coverage \cite{AsadOverview}. According to Statista, there were 1,468 operational LEO satellites in September 2019. In the past five years, this number has surged by approximately 467 \%, reaching over 8,400 active LEO satellites in September 2024. This number is expected to grow further to 100,000 by 2030. 
The past years have delivered immense progress in satellite broadband connectivity, with Starlink among the most prominent examples. The Starlink mega-constellation comprises over 6,912 LEO satellites as of January 2025. Starlink uses proprietary technology to deliver high-speed and low-latency Internet to residential users, and businesses inland and on the ocean as well as in-motion connectivity for vehicles, vessels, and aircrafts \cite{Garcia,Dominic_2024,Beckman,Zhao}.

For a successful and stable Internet connection, an unobstructed field-of-view (FoV), which leads to a clear line-of-sight (LOS) between the terrestrial terminal and Earth-orbiting satellites, is essential. Any obstruction that blocks LOS and reduces effective FoV can lower throughput, increase round-trip time (RTT), and cause frequent connection outages \cite{Sami,Dominic_2024}. These disruptions also negatively impact website browsing, video streaming, video calls, and online gaming applications \cite{Zhao,Michel,Kassem}. Over the past few years, several studies have examined Starlink's performance in the rural and urban areas for stationary and mobile scenarios~\cite{Michel, Kassem, Sami, Dominic_2024}. However, the majority of these studies use the terminals designed for stationary residential connectivity to conduct in-motion performance analysis. For instance, in~\cite{Bin,Sami,Melisa}, residential terminals are used to examine the Starlink performance in motion. However, Starlink advises against using these terminals while in motion.

In December 2022, Starlink introduced the Flat High-Performance (FHP) terminal, classified as an Earth Station in Motion (ESIM), for use within the United States. FHP was launch in Europe in April 2023 \cite{Dominic_2024}. FHP terminal is designed to offer broadband connectivity to moving vehicles and vessels. As of now, a limited number of studies have examined the performance of the FHP terminal while in motion \cite{Dominic_2024, Bin, Beckman}, only \cite{Beckman} evaluates Starlink's performance in Northern Europe. However, it primarily focuses on investigating the downlink performance and latency. 
The number of Starlink measurement studies in Northern Europe is limited, and satellite communication performance near the poles typically differs from that in regions closer to the equator. For example, the Starlink constellation is sparse, and the Global navigation satellite system has poor functionality in Northern Europe compared to the regions near the equator.

This motivates us to look at the Starlink's characteristics in Northern Europe. This paper investigates the capabilities of the FHP terminal more extensively compared to earlier FHP studies \cite{Dominic_2024, Bin, Beckman}. Our main contributions  are as follows:
\begin{itemize}
 \item We conduct an up-to-date state-of-the-art review of the Starlink system. Specifically, we accumulate the information from our experimental measurements, state-of-the-art publications \cite{Garcia,Rodrigo,Dominic_2024,Sami,Beckman,Melisa,Bin,Zhao,Michel,Kassem,Marko_2}, and technical specifications \cite{StarlinkAccessories}.
  \item  To the best of our knowledge, this is the first paper to report the FHP terminal uplink and downlink throughput in Northern Europe. We conduct stationary and in-motion measurements to examine the key performance indicators, including RTT, routing strategy, and bi-directional throughput. We compare the RTT and routing strategies of Starlink with those of the terrestrial network.
  \item We assess network performance from a client in Northern Finland to different servers located in Northern Europe, Western Europe, and the United States. This enables us to examine and compare Starlink's performance across different geographical regions.
   \item We present throughput and latency in motion as a function of geospatial data and vehicle speed. This allow us to observe the impact of the roadside blockages and variable vehicle speed on the communication performance. 
  \item  To facilitate further research, we have made our dataset publicly accessible, along with the relevant script codes for data analysis, on GitHub [To be added later]. 
\end{itemize}

\begin{table*}[t!]
    \centering
    \caption{Starlink terminal and key specifications as of January 14, 2025.}
    \begin{tabularx}{\textwidth}{|l|>{\centering\arraybackslash}X|>{\centering\arraybackslash}X|>{\centering\arraybackslash}X|>{\centering\arraybackslash}X|>{\centering\arraybackslash}X|}
        \hline
        \textbf{Functions} & \textbf{Standard} & \textbf{Standard} & \textbf{High} & \textbf{Flat High} & \textbf{Mini} \\
        & \textbf{} & \textbf{Actuated} & \textbf{Performance} & \textbf{ Performance} & \textbf{} \\
        \hline
      \textbf{Antenna} & EPA & EPA & EPA & EPA & EPA \\
       \textbf{Dimension (mm $\times$ mm)} & 594 $\times$ 383 & 513 $\times$ 303 & 575 $\times$ 511& 575 $\times$ 511 & 298.5 $\times$ 259 \\
       \textbf{Field of View} & 110$^\circ$ & 100$^\circ$ & 140$^\circ$ & 140$^\circ$ & 110$^\circ$ \\
        \textbf{Orientation} & SAMO & MSO & MSO & Fixed & SAMO\\
      \textbf{Snow Melt Capability (mm/h)} & 40&  40 & 75  & 75  &  25 \\    
\textbf{Average Power Usage (\SI{}{\watt})} & 75-100& 50-75 &  110-150& 110-150 &  25-40\\
    \textbf{Wind Rating (km/h)} & 96+ & 80+ & 80+ & 280+ & 96+ \\
\textbf{In-motion capability} &  Permitted & up to
16 km/h & up to
16 km/h & Recommended &  Permitted\\
\textbf{Terminal weight (\SI{}{\kilogram})} & 2.9 & 2.9  & 6.9  & 5.9 & 1.1\\
        \hline
    \end{tabularx}
    \label{tab:TabI}

    \vspace{-5pt}
\end{table*}

The rest of this paper is organized as follows. Section~\ref{sec:sec2} presents the Starlink specifications and discusses the state-of-the-art measurement studies. In Section~\ref{sec:sec3} and Section~\ref{sec:sec4}, we present the measurement setup and collected dataset, respectively.  Section~\ref{sec:sec5} discusses the measurement results and the lessons learned. Finally, Section~\ref{sec:sec6} concludes this paper with final remarks.

\vspace{-4pt}

\section{Starlink: Terminals and Service Plans}
\label{sec:sec2}

\vspace{-4pt}
Starlink offers Internet connectivity for residential users, roaming (mobile) option for recreational vehicles (RVs) and campers as well as in-motion connectivity for the vehicles and vessels \cite{Dominic_2024,Marko, Marko_2,Garcia}. For each use case, Starlink recommends a specific ground terminal with a tailored service plan to ensure optimal performance. A summary of the available terminals and service plans is presented below.

\vspace{-4pt}
\subsection{Terminals}
Presently, Starlink offers five ground terminal options, all use an Electronic Phased Array (EPA) antenna and can operate in temperate from -30$^\circ$C to  50$^\circ$. Table~\ref{tab:TabI} compares the specifications of commercially available terminals. Notably, Starlink recommends only the FHP terminal for in-motion use. We discuss below the key characteristics. 
\subsubsection{Standard} This terminal offers 110$^\circ$ FoV and is designed for residential use to support everyday Internet applications e.g., video streaming, video calls, and online gaming. The lightweight with \SI{2.9}{\kilogram}  makes it easier to carry and a suitable connectivity option for RVs, campers, and travelers.  The Standard terminal features Software-Assisted Manual Orienting (SAMO) which requires manual alignment with the help of the Starlink smartphone application.
\subsubsection{Standard Actuated} Similar to the Standard terminal, the Standard Actuated terminal is designed for residential users. However, it has 10$^\circ$ degree smaller FoV and consumes \SI{25}{\watt} less average power than the Standard terminal. Another key difference between these terminals is their alignment and orientation adjustment capabilities. The Standard Actuated terminal features  Motorized Self-Orienting (MSO) alignment which automatically points the terminal toward an optimal direction in the sky.  
\subsubsection{High Performance} This terminal is designed to offer better performance for supporting high-demand users and enterprise connectivity. It offers 140$^\circ$ FoV and better resilience in extreme environments due to its higher snow-melting capabilities, handling up to 75 mm/h compared to 40 mm/h for Standard and Standard Actuated terminals. On the downside, it consumes more power.
\subsubsection{Flat High Performance} The flat surface makes the FHP terminal more resilient to wind with a tolerance of 280 km/h. FHP terminal is a type of ESIM and can be installed on the rooftops of vehicles \cite{Dominic_2024,Beckman,Bin} and vessels as well as could be mounted on a pole for residential users \cite{StarlinkAccessories}. Similar to High Performance terminal, the FHP terminal offers 140$^\circ$ field of view making it visible to more satellites.  Due to the fixed installation, this terminal neither supports SAMO nor MSO alignments correction capability \cite{StarlinkAccessories}.
\subsubsection{Mini} Most recently, Mini was introduced in July 2024. As the name suggests, this terminal is compact, lightweight, and portable.  Mini can fit in a backpack and includes an integrated Wi-Fi 5 router. It is capable of providing high-speed, low-latency Internet, however, the throughput is lower compared to other terminals. Similar to the Standard terminal, Starlink permits the Mini for in-motion connectivity but does not recommend it. However, Starlink recommends Mini for roaming from one location to another.
\begin{table*}[t!]
    \centering
    \caption{Starlink Service Plans, Recommended hardware terminals, average performance and use cases as of September 2, 2024.}
    \begin{tabular}{|l|c|c|c|c|c|c|c|c|c|}
        \hline
        \textbf{Service plan} & \textbf{Standard Data} & \textbf{Cost} & \textbf{In-motion} & \textbf{Recommended}& \textbf{Recommended}& \textbf{Downlink} & \textbf{Uplink} & \textbf{Latency}\\
        \textbf{} & \textbf{(Priority Data)} & \textbf{} & \textbf{} & \textbf{hardware}& \textbf{use case}& \textbf{Mbps}& \textbf{Mbps}& \textbf{ms}\\
        \hline
       
        \textbf{Standard (fixed)} & Unlimited & 50 € & $\Cross$ & Standard &  Households& 25-100 & 5-10  & 25-60\\
       \textbf{Priority (fixed)} & Unlimited (40 GB) & 70 € & $\Cross$ & FHP &  Businesses& 40-220 & 8-25  & 25-60 \\
        \textbf{Mobile (mobility)} & Unlimited  &59 € &  less 16 km/h &  Mini &  Campers&5-50 &  2-10 & $<$99\\
        \textbf{Mobile priority (mobility)} & Unlimited (50 GB)  &296 € & \checkmark & FHP & In-motion use &40-220  & 8-25  & $<$99\\
        \hline
    \end{tabular}
    \label{tab:TabII}

    \vspace{-5pt}
\end{table*}

\subsection{Service Plans}
\label{ServicePlans}

Starlink offers both fixed and mobility service plans, with and without priority options, designed to meet the connectivity needs of stationary residential and mobile users as well as businesses. Table~\ref{tab:TabII} compares the available four service plans.

Without priority, both Standard (fixed) and Mobile (mobility) service plans offer downlink throughput 25-100 Mbps and 5-50 Mbps, respectively and uplink throughput 5-10 Mbps and 2-10 Mbps, respectively. According to Fair Use Policy~\footnote{Starlink’s Fair Use Policy (https://www.starlink.com/legal/documents/DOC-1469-65206-75)}, Starlink distributes the data among active users fairly and equitably. During network congestion, a user with a priority-enabled service plan will be given preference to experience faster and more consistent throughput than a non-priority user.

With the priority capability, both Priority (fixed) and Mobile Priority (mobility) service plans promise to deliver downlink throughput 40-220 Mbps and an uplink throughput 8-25 Mbps. According to Starlink specifications~\footnote{Starlink Specifications (https://www.starlink.com/legal/documents/DOC-1400-28829-70)}, the Standard (fixed) and Priority (fixed) service plan have latency between 25 ms to 60 ms. For Mobile (mobility) and Mobile Priority (mobility) service plans, latency is less than 99 ms.

\begin{table*}[t]
    \centering
    \caption{List of the relevant reviewed papers, including the used terminal, protocol, link direction, and measurement region.}
 \begin{tabularx}{\textwidth}{|l|c|c|c|c|>{\centering\arraybackslash}X|c|c|}
        \hline
        \textbf{Ref.} & \textbf{Publication}& \textbf{FHP} & \textbf{Communication} & \textbf{Measurement} & \textbf{Link} & \textbf{Region} &  \textbf{In-motion}\\
        \textbf{} & \textbf{year}& \textbf{included} & \textbf{protocol} & \textbf{tool} & \textbf{analysis} & \textbf{country} &  \textbf{included}\\
        \hline
         \textbf{\cite{Sami}} & 2022&  $\Cross$& TCP/UDP & \verb|iperf3|  & Downlink, Uplink & Canada & \checkmark\\
        \textbf{\cite{Michel}} &2022& $\Cross$ & TCP/QUIC & Ookla SpeedTest & Downlink, Uplink & Belgium & $\Cross$\\
       \textbf{\cite{Kassem}} & 2022 & $\Cross$ & TCP & \verb|iperf3|, Libretest& Downlink, Uplink & UK,
USA, EU and Australia &  $\Cross$\\
 
        \textbf{\cite{WetLinks}} & 2024 & $\Cross$ &  UDP& \verb|iperf3|& Downlink, Uplink &  Germany, Netherlands& $\Cross$\\
        \textbf{\cite{Bin}}  & 2023 & \checkmark &  TCP/MPTCP/UDP& \verb|iperf3| & Downlink, Uplink &  USA& \checkmark\\
       \textbf{\cite{Beckman}} & 2023 & \checkmark &  TCP& Router logs & Downlink &  Sweden& \checkmark\\
        \textbf{\cite{Dominic_2024}} & 2024 & \checkmark &  UDP& \verb|iperf3|& Downlink, Uplink &  Germany& \checkmark\\
        \textbf{Our work} & 2025 & \checkmark &  TCP& \verb|iperf3|& Downlink, Uplink &  Finland& \checkmark\\
        \hline 
    \end{tabularx}
    \label{tab:TabIII}

    \vspace{-10pt}
\end{table*}
\subsection{Related Works}
Recent years have witnessed several measurement campaigns which investigate the Starlink's performance characteristics \cite{Dominic_2024,Sami,Beckman,Melisa,Bin,Zhao,Michel,Kassem}. Table \ref{tab:TabIII} lists the selected related works, publication year, and information about the experimental setup and the geographical region. In \cite{Sami}, the authors conducted measurements from urban and remote areas of Canada using Starlink Generation-1 (Gen-1, round-shaped) and Generation-2 \footnote{It appears from the figures (see Fig. 13 (a) in \cite{Sami}) that Gen-2 rectangular dish was Standard Actuated terminal.} (Gen-2, rectangular) ground terminals. The results confirm that Starlink has the potential to provide ubiquitous Internet connectivity. Both Gen-1 and Gen-2 terminals offered comparable performance. Notably, it was observed that Starlink's throughput and latency fluctuate over time compared to terrestrial networks, and the connection experiences frequent outages. Additionally, in-motion tests were conducted for 30 minute period, with vehicle speeds ranging from 40 km/h to 70 km/h, revealed very unstable performance and frequent outages. It is worth emphasizing that these mobility tests were conducted using a Gen-1 terminal, which was primarily designed for stationary use. To the best of our knowledge, the Gen-1 terminal is now obsolete, and Starlink no longer advertises or sells this round-shaped model. 

The work in \cite{Michel} presents measurements conducted in Western Europe, evaluating Starlink's  latency, packet loss rate and throughput using Transmission Control Protocol 
 (TCP) and Quick UDP Internet Connection (QUIC). The authors in~\cite{Kassem} examine website browser performance using data from eighteen Starlink user terminals across the UK, USA, EU, and Australia. Additionally, the work in \cite{Zhao} discusses measurements conducted in urban and rural areas of the USA. Collectively, the studies \cite{Michel, Kassem, Zhao,Sami} reveal that Starlink can support demanding applications including live streaming, video conferencing, and cloud gaming. In \cite{WetLinks}, stationary measurements were conducted to collect throughput data using the UDP and to examine the impact of weather conditions on performance in Germany and the Netherlands.

As of today, only a few studies examine the Starlink performance when in-motion. In \cite{Bin,Sami,Melisa}, residential terminals are used for in-motion measurements. However, these terminals are not designed for mobility. Starlink advises against in-motion use of these terminals due to the risk of equipment falling onto the road and serious road accidents\footnote{https://support.starlink.com/topic?category=26}.

The works in \cite{Bin} and \cite{Beckman} compare the Starlink FHP terminal and cellular networks in motion measurements conducted in the USA and Sweden, respectively. The comparison indicates that, on average, Starlink outperforms cellular networks. However, Starlink experiences more frequent connection outages. Starlink measurements in Sweden within the Arctic Circle solely focus on the downlink analysis \cite{Beckman}. More recently in \cite{Dominic_2024}, the FHP terminal download and upload UDP throughput results are reported for stationary and in-motion scenarios in Germany that locates in Central Europe. It was observed that the Starlink in-motion performance is significantly worse than stationary performance. Specifically, the uplink and downlink performance decreases each by nearly 10 \%. Additionally, a lower throughput was observed in urban areas due to LOS obstruction by buildings. However, the number of samples for stationary and in-motion measurements were not equal.

Based on the literature review, there is a clear need to conduct a more thorough bi-directional measurement campaign to study the performance of the new FHP terminal in Northern Europe for both stationary and in-motion use. Our paper addresses this shortcoming with measurements to better understand its throughput and usefulness for applications.

\if{0}
\subsection{Terminal Obstruction Map}
The Starlink app enables users to scan the sky, identify potential blockages, and find a suitable unobstructed location to install the terminal. Within one week of connection with overhead satellites, Starlink generates an obstructions map identifying the possible blockages (e.g., trees, poles, and buildings). The accuracy of this obstruction map improves over time as additional data is collected. To optimize performance, Starlink uses this data and priorities to communicate with satellites located in unobstructed areas of the sky whenever feasible. However, in the case of in-motion use, the communication channel continuously changes due to mobility, making it challenging to benefit from the obstruction map.
\fi

\section{Experimental setup}
\label{sec:sec3}

\vspace{-1pt}
\subsection{Hardware}
\begin{figure}[t!]
\centerline{\includegraphics*[width=0.48\textwidth]{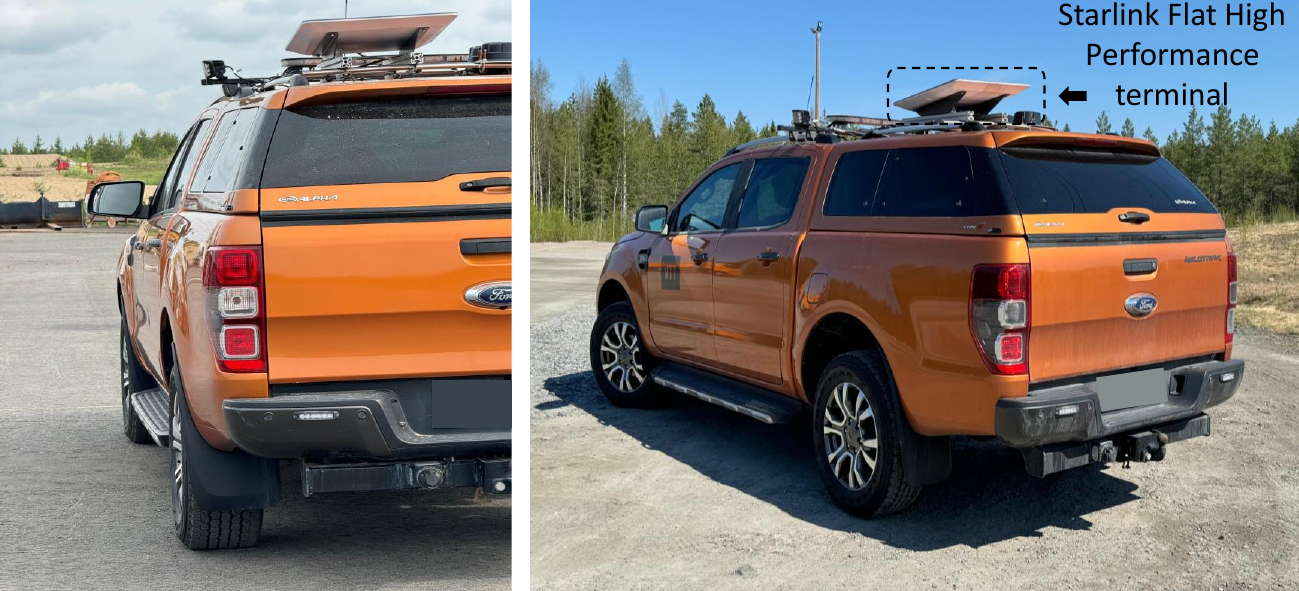}}
\caption{Ford Ranger vehicle with installed Flat High Performance terminal.}
\label{fig:fig1}

\vspace{-10pt}
\end{figure}
\begin{figure}[t!]
\centerline{\includegraphics*[width=0.5\textwidth]{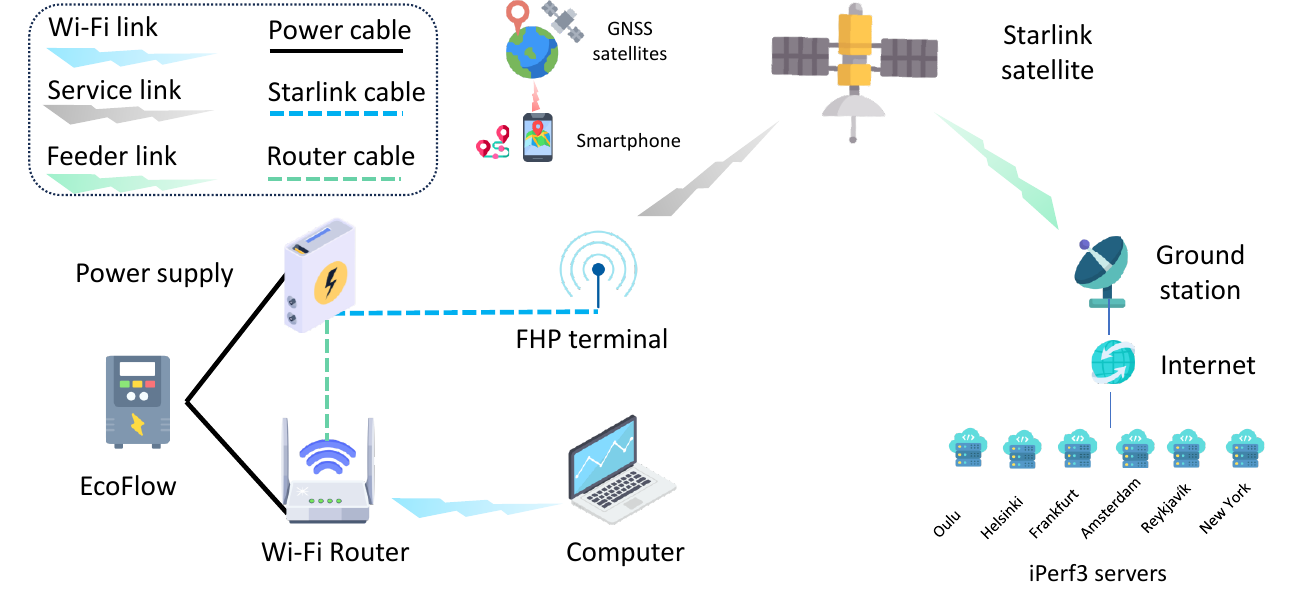}}
\caption{Experimental setup illustration.}
\label{fig:fig2}

\vspace{-10pt}
\end{figure}

We installed the Starlink's FHP terminal on the rooftop of the a Ford Ranger vehicle. Specifically, the terminal was connected to the Wedge Mount~\cite{StarlinkAccessories} which was further installed on the Ford Ranger's roof racks as shown in Fig.~\ref{fig:fig1}. The Starlink Wedge Mount comes with 8$^\circ$ tilt to facilitate water runoff during the rain. Next, we connected the terminal with Starlink's power supply module using the~Starlink Cable. The power supply module was further connected to Starlink's Wi-Fi router using Router Cable. We used EcoFlow portable power station which is a battery-powered generator to provide the required electric input to the installed setup. The measurements were conducted using a laptop featuring a 64-bit operating system. The laptop's processor was an Intel(R) Core(TM) i5-6200U CPU operating at 2.30 GHz, and it featured an Intel(R) Dual Band Wireless-AC 8260 Wi-Fi adapter. The system was running Windows 10 Enterprise edition. We connected laptop with  Starlink's Wi-Fi router using Wi-Fi connection. Fig.~\ref{fig:fig2} illustrates the components of our experimental setup. To facilitate the throughput analysis when in-motion and understand the impact of roadside obstacles, we used a smartphone application called GPS Logger\footnote{GPS Logger (https://github.com/BasicAirData/GPSLogger)}. This application stores the vehicle location and vehicle speed every second into a database.

\subsection{Software}
To be consistent with earlier studies~\cite{Sami,Dominic_2024,Bin,WetLinks,Kassem}, we used the following three tools for measurements.
\subsubsection{iPerf3} 
This tool measures the maximum achievable throughput of Internet protocol (IP) networks. \verb|iperf3| supports measurements in both uplink and downlink directions, client-to-server and server-to-client, respectively. By default, \verb|iperf3| uses the CP, but also provides a UDP option. Additionally, \verb|iperf3| allows for the transmission of multiple parallel streams simultaneously, facilitating the assessment of the network's maximum achievable performance. In this paper, we used TCP for both directions with the Windows default congestion control algorithm.
\if{0}
Specifically, we used \verb|iperf3| in Windows Command Prompt as 


\begin{Verbatim}[fontsize=\footnotesize]
Uplink: iperf3 -c <IP address> -p 5201 -P 10
Downlink: iperf3 -c <IP address> -p 5201 -R -P 10
\end{Verbatim}
where \verb|<IP address>| of \verb|iperf3| servers are discussed in Section \ref{subsec:server}, \verb|-P 10| generates ten parallel streams to fully test the link, and \verb|-R| sets communication mode to downlink.
\fi
\subsubsection{Ping} We used \verb|ping| to measure RTT. To be consistent with earlier studies \cite{WetLinks,Bin}, in this paper, the term RTT is understood as two-way latency between the client and the server. For consistency, we used the same \verb|IP address| discussed in Section \ref{subsec:server} for both \verb|ping| and \verb|iperf3| measurements. Additionally, we used the Windows default \verb|ping| setting,  where a single \verb|ping| test sends uplink packets (each 32 bytes long) to a server and expects to receive the same number of packets as a reply from a server in the downlink.
\subsubsection{traceroute} We used \verb|tracert| to understand how the uplink data travels from the client to the server. It helps to understand the following questions: (i) What is total number of hops between the client and server? (ii) Where is the nearest ground station? (iii) How does data travel from the ground station to the target server? 
\if{0}
\begin{figure}[t!]
\centerline{\includegraphics*[width=0.5\textwidth]{ServerLocations3.eps}}
\caption{Geographical locations of investigated iPerf3 servers.}
\label{fig:fig3}
\end{figure}
\fi
\subsection{Client and Server Locations}
\label{clientLocation}
The client and server locations in the test are as follows.
\subsubsection{Client location (stationary)}
We conducted stationary measurements at four different parking areas by parking the vehicle carrying the Starlink FHP terminal. The measurement sites in Oulu, Finland include (i) the VTT Technical Research Center Ltd; (ii) the Botanical Garden of the University of Oulu; (iii) Auran Maja, a cross-country skiing resort located approximately \SI{10}{\kilo\meter} from the Oulu city; and (iv) OuluZone, a multi-purpose area located \SI{35}{\kilo\meter} from the Oulu, commonly used for radio test measurements. There were no blockages in the surrounding area at (i), (ii), and (iii). However, the vehicle and FHP terminal were positioned facing roadside trees during the measurements at location (ii).

\subsubsection{Client locations (in-motion)}
\label{Mobilitylocation} When in-motion, the communications channel condition continuously changes due to both vehicle and satellite mobility, making communication further challenging. To examine in-motion performance, we selected a \SI{6}{\kilo\meter} route beginning and ending at VTT, Oulu. The route involves entering the European motorway E8 at Exit 12, proceeding to Exit 11, making a return, and then exiting the E8 again at Exit 12 upon returning to VTT. The traveled path on E8 has a maximum speed limit of 100 km/h.

\subsubsection{Servers locations}
\label{subsec:server}
To investigate how Starlink data travels between the client and server and to assess the impact of different geographical locations on communications performance, we selected six \verb|iperf3| servers located at diverse geographical locations as shown in Table \ref{tab:tableIV}. Thanks to Hostkey\footnote{Hostkey B.V. Speedtest (https://speedtest.hostkey.com/)}, it offers \verb|iperf3| servers at five locations including (i) Helsinki, Finland; (ii) Amsterdam, Netherlands; (iii) Frankfurt, Germany; (iv) Reykjavík, Iceland; and (v) New York, United States. Notably, Starlink offers 100 \% coverage to these cities\footnote{Starlink Availability Map (\url{https://www.starlink.com/map})}.
\begin{table}[!t]
\caption{The IP addresses of examined iPerf3 servers\label{tab:tableIV}}
\centering
\begin{tabular}{|l|c|c|c|}
\hline
No. & Server location & Server IP&Owner\\
\hline
1& Helsinki, Finland & spd-fisrv.hostkey.com & Hostkey\\
2.& Amsterdam, Netherlands& spd-nlsrv.hostkey.com & Hostkey\\
3.& Reykjavík, Iceland &spd-icsrv.hostkey.com & Hostkey\\
4.& Frankfurt, Germany& spd-desrv.hostkey.com & Hostkey\\
5.& New York, United States& spd-uswb.hostkey.com & Hostkey\\
6.& Oulu, Finland & Anonymous & VTT\\
\hline
\end{tabular}

\vspace{-10pt}
\end{table}

\if{0}
\begin{Verbatim}[fontsize=\footnotesize]
1. Helsinki, Finland spd-fisrv.hostkey.com
2. Amsterdam, Netherlands spd-nlsrv.hostkey.com
3. Reykjavik, Iceland spd-icsrv.hostkey.com
4. Frankfurt, Germany spd-desrv.hostkey.com
5. New York, United States spd-uswb.hostkey.com
\end{Verbatim}
\fi
Additionally, we set up a sixth \verb|iperf3| server in the same city as the client, Oulu, Finland which was running in a computer with \verb|Ubuntu 22.02| operating system. The computer was in VTT’s test network\footnote{5G and 6G test network environment by VTT (\url{https://www.vttresearch.com/en/5g-and-6g-test-network-environment})}, and it was connected to the Internet using a 1 Gbps Ethernet interface. The computer has a public IP address and TCP port  was open to the Internet during the test. The \verb|iperf3| version was 3.17 and we used a default configuration including TCP congestion control. Due to security reasons, we are not disclosing the IP address. 

\section{Measurements and Data Collection: Overview}
\label{sec:sec4}
To compare Starlink's performance in both stationary and in-motion scenarios, this paper reports results from four days including May 24, June 7, June 26, and June 27, 2024. These dates were chosen due to the similarity in weather conditions during the measurements and to make a fair comparison. Notably, the sky was mostly clear with no rain, ensuring consistency in weather conditions. In the future, we plan to conduct measurements under challenging weather conditions.

\vspace{-4pt}
\subsection{Dataset and Service Plans}
Our Starlink FHP terminal has 50 Gigabit/month Mobile Priority data subscription which was fully used during the measurements reported in Fig. \ref{fig:fig7}. The analysis and measurements reported in Fig. \ref{fig:fig4}, Fig. \ref{fig:fig8} and Fig. \ref{fig:fig9} were performed when the Mobile Priority data was already consumed. According to Starlink’s Fair Use Policy\footnote{Starlink Fair Use Policy (\url{https://www.starlink.com/legal/documents/DOC-1469-65206-75})}, the user who has already consumed the allocated Mobile Priority data will be offered unlimited Mobile data for the rest of the month. For comparison, we also conducted \verb|tracert| and \verb|ping| tests using a 1 Gbps wired Internet connection.
\subsection{Stationary Measurements} In Fig. \ref{fig:fig4} \verb|ping| measurements consist of 774 samples for each server for Starlink. The terrestrial network measurements comprise the same number of samples. This gives us a total of 9,288 samples for both networks. In Fig. \ref{fig:fig7} Mobile Priority measurements were conducted at VTT, the Botanical Garden, and Auran Maja on May 24, 2024, while additional measurements at OuluZone were carried out on June 7, 2024. For each location and in each direction downlink/uplink, we collected 599 \verb|iperf3| measurement samples, resulting in a total of  4,792 measurement samples (averaged over ten \verb|iperf3| parallel streams) for all four locations.
For the results in Fig. \ref{fig:fig8}, we conducted measurements on two consecutive days, June 26 and June 27, 2024. During the measurements, the vehicle was parked in the VTT parking lot, and we collected a total of 12,914 \verb|iperf3| measurement samples for uplink and downlink for all servers, except the one in New York. Specifically, we conducted an \verb|iperf3| test for each server, running separate uplink and downlink tests for 300 seconds. However, there were inconsistencies in the measurements sample length for each sever due to two reasons:
\begin{itemize}
  \item Server availability: We repeated our measurements five times for each server listed in Table \ref{tab:tableIV}. However, some servers were unavailable during the test due to being occupied by other clients. For instance, the Hostkey server in the USA was consistently busy and returned the error message \textit{iperf3: error - the server is busy running a test. try again later} during all five attempts.
  \item Measurement interruptions: For the available servers, some measurements were interrupted unexpectedly with an error message \textit{iperf3: error - control socket has closed unexpectedly} indicating interruption from the server side.
\end{itemize}
 
From our five tests, we collected downlink \verb|iperf3| samples as 1,500, 1,500, 1,061, 1,212, and 1,200 for servers located in Oulu, Helsinki, Frankfurt, Amsterdam, and Reykjavík, respectively. The uplink dataset has 1,500, 1,222, 1,019, 1,500, and 1,200 samples for the same servers. This resulted in a total of 12,914 samples for the results presented in Fig. \ref{fig:fig8}.

\subsection{In-motion measurements}
For the in-motion measurements, we drove the vehicle five times on the path defined in Section \ref{Mobilitylocation}. Four of these drive tests were conducted to collect \verb|iperf3| data, comprising two drive tests for uplink and two for downlink. The fifth drive test was performed to collect \verb|ping| samples. We collected  2068 in-motion samples, on average 830 \verb|iperf3| samples for uplink and downlink respectively, and 396 samples for \verb|ping|. During these drive tests, the average vehicle speed was around 50 km/h and the highest speed was 100 km/h.

\section{Measurement results}
\label{sec:sec5}
\subsection{Routing and Round-trip Time}
\begin{figure}[t!]
\centerline{\includegraphics*[width=0.5\textwidth]{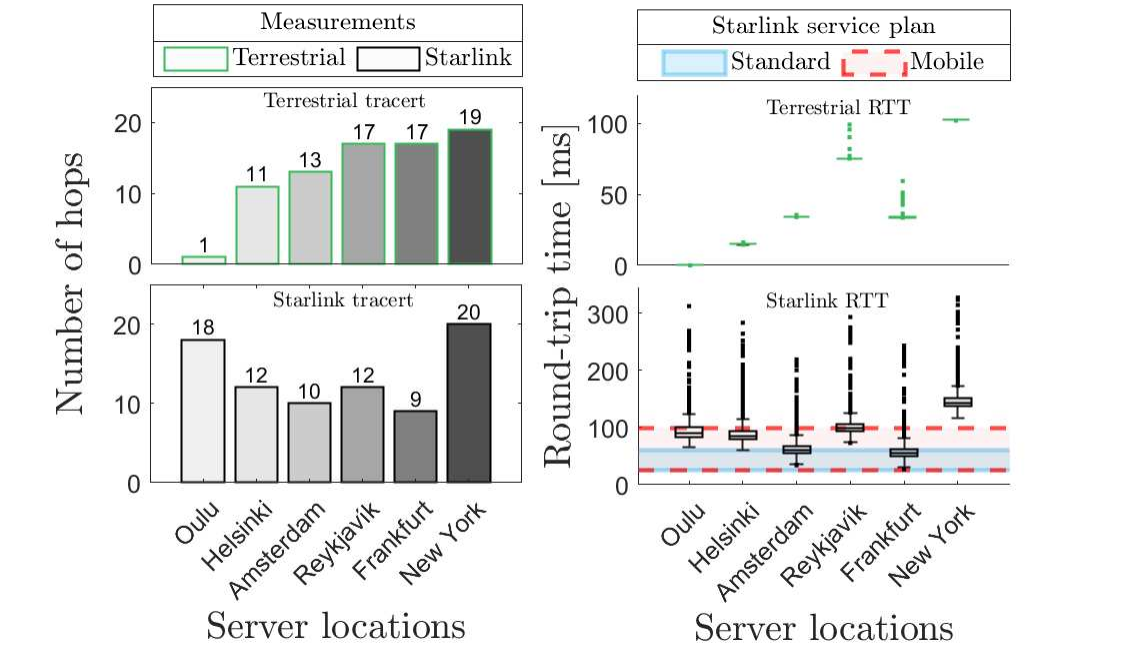}}
\caption{Comparison of the number of hops (left) over the route and round-trip time (right) for terrestrial (top) and Starlink (bottom) networks from Oulu to six iperf3 servers located in (i) Oulu; (ii) Helsinki; (iii) Amsterdam; (iv) Reykjavík; (v) Frankfurt and (vi) New York.}
\label{fig:fig4}

\vspace{-10pt}
\end{figure}
Fig. \ref{fig:fig4} compares the number of hops  (left) over the client to server route obtained from \verb|tracert| and \verb|ping| RTT (right) for terrestrial (top) and Starlink (bottom) networks from Oulu to six iperf3 servers as discussed in Section \ref{subsec:server} and Table~\ref{tab:tableIV}. When using the terrestrial network, there was only 1 hop and a relatively low RTT for the server in Oulu. This was because the client and server were installed on the same computer. When using Starlink, the server in Oulu experiences 18 hops. For the server in New York, both terrestrial and Starlink connection gives a comparable number of hops 19 and 20, respectively. 

With Starlink, the servers located in Oulu and New York have almost double hops compared to Frankfurt and Amsterdam which have 9 and 10 hops, respectively. However, this is not the case with terrestrial networks, where the servers located in Helsinki, Amsterdam, Reykjavík, and Frankfurt have 11, 13, 17, 17, and 19 hops, respectively.

Our \verb|tracert| analysis reveals that Starlink terminal in Oulu uses ground stations in Germany/Belgium. Therefore, the servers in Frankfurt and Amsterdam have a low number of hops compared to other locations. For all the servers including the one in New York, the traffic always passes through the IP address belonging to the USA as indicated in \cite{Marko_2}. The data traffic started using the terrestrial network after being received by a ground station in Belgium or Germany.  In this paper, we report Starlink \verb|tracert| results for a single iteration on July 24, 2024. Terrestrial \verb|tracert| results are for three iteration conducted on August 24, 2024. The number of routes may fluctuate slightly over time, potentially during the network congestion and the chosen service plan.

The right-hand side of the Fig. \ref{fig:fig4} illustrates boxchart\footnote{Box chart (\url{https://se.mathworks.com/help/matlab/ref/boxchart.html})} representing the median RTT obtained from \verb|ping| measurements across these six locations. Each boxchart shows the following information (a) the median which is the line inside of each box, (a) the lower and upper quartiles represented by the top and bottom edges of each box (c) outliers illustrated using dots ``.'', and (d) the minimum and maximum values excluding outliers indicated by the whiskers lines extend above and below each box. The \verb|ping| median results indicate that New York exhibits the highest median RTT of 103 ms and 143 ms for terrestrial and Starlink connections, respectively. This is followed by Raykjavik having median RTT of 75 ms and 99 ms for the terrestrial and Starlink connections, respectively. With Starlink connection, Frankfurt and Amsterdam demonstrate the lowest RTT with a median of 55 ms and 60 ms, respectively. This median RTT reduces to around 33 ms for the terrestrial network. Similarly, Oulu's terrestrial median RTT is 0.028~ms which increases to 90 ms for Starlink because the data goes from Oulu to  Frankfurt using satellite and returns from Frankfurt to Oulu using the terrestrial network. Compared to terrestrial connection, Starlink increases Helsinki's median RTT from 15 ms to 85 ms. In Fig. \ref{fig:fig4} blue and red areas are references for Starlink's expected performances in Standard and Mobile service plans, respectively. Except for the server in New York, the RTT for all other servers remained within RTT range supported by the Mobile service plan.

\begin{figure}[t]
\centerline{\includegraphics*[width=0.5\textwidth]{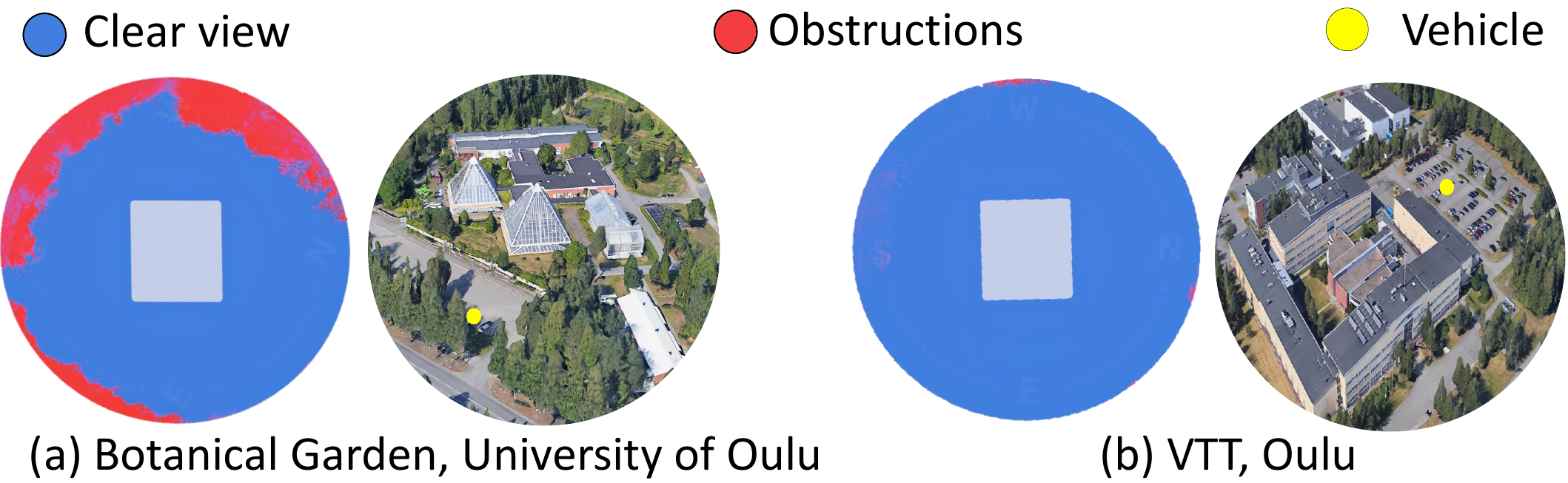}}
\caption{Obstruction Map of the two test locations (a) Botanical Garden and (b) VTT where blue color shows the clear LOS and red dots indicate blockages.}
\label{fig:fig6}

\vspace{-3pt}
\end{figure}

\begin{figure}[t]
\centerline{\includegraphics*[width=0.5\textwidth]{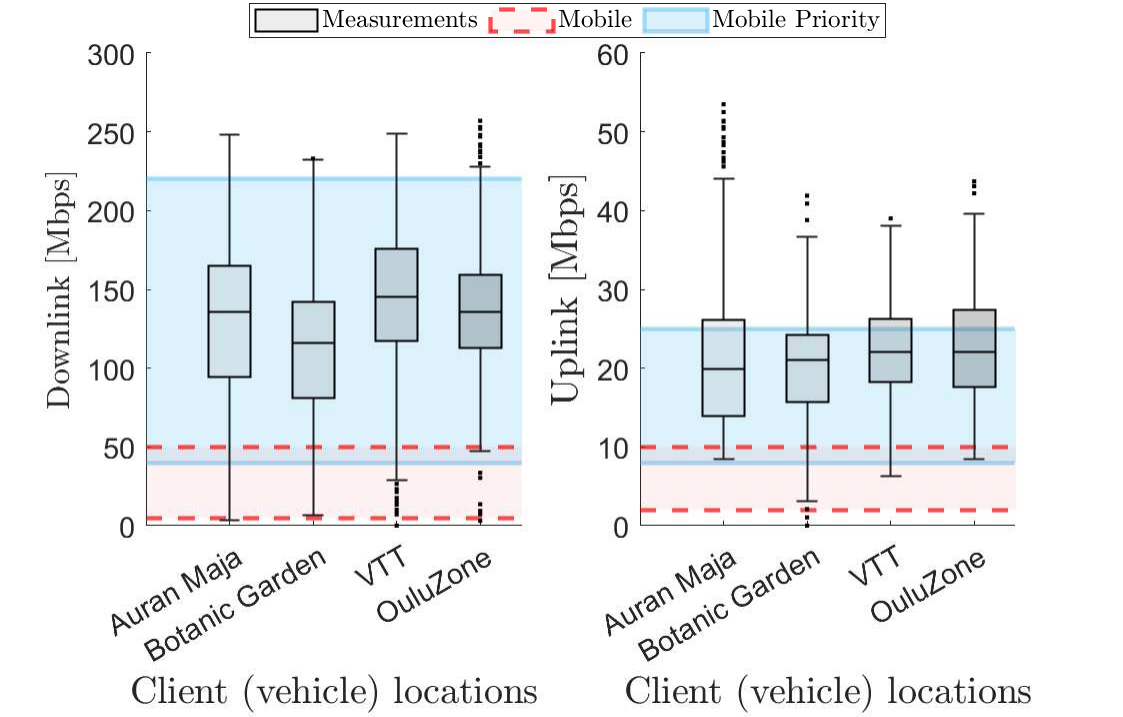}}
\caption{Starlink downlink and uplink throughput from four different test locations in Oulu to iperf3 servers in Finland.}
\label{fig:fig7}

\vspace{-10pt}
\end{figure}

\vspace{-4pt}
\subsection{Throughput}
The Starlink smartphone application allows users to scan the sky to find an unobstructed area to install the terminal for uninterrupted connectivity. To visualize the possible obstructions at the client location discussed in \ref{clientLocation}, we used the Starlink smartphone application to generate an Obstruction Map. Fig. \ref{fig:fig6} illustrates the Obstruction Map for two test locations Botanical Garden and VTT. In Fig. \ref{fig:fig6} Obstruction Map, blue regions indicate areas with unobstructed FoV, while red areas show the obstructed FoV due to roadside trees.

\begin{figure*}[t!]
    \centering
    \begin{minipage}[t]{0.497\textwidth}
        \centering
        \includegraphics[width=\textwidth]{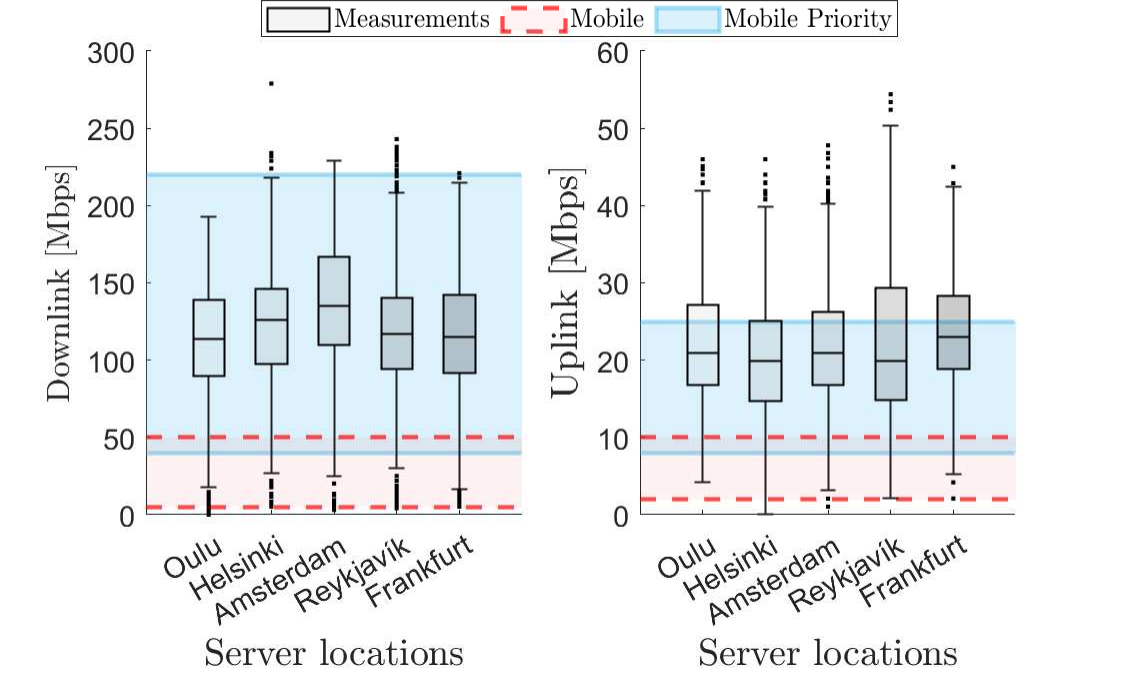}
        \caption{Starlink downlink and uplink Internet speed from VTT's parking, Oulu to (i) Oulu, (ii) Helsinki, (iii) Amsterdam, (iv) Frankfurt, and (v) Reykjavík.}
        \label{fig:fig8}
    \end{minipage}
    \hfill
    \begin{minipage}[t]{0.497\textwidth}
        \centering
        \includegraphics[width=\textwidth]{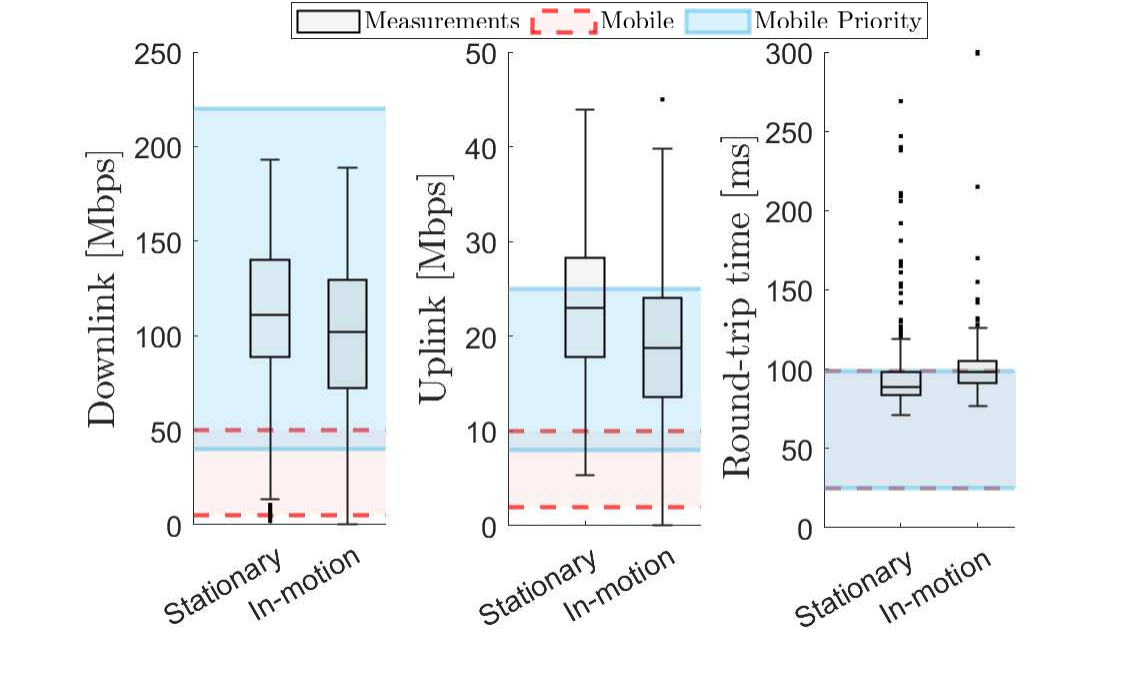}
        \caption{Comparison of Starlink downlink and uplink throughput and RTT for stationary and in-motion scenarios.}
        \label{fig:fig10}
    \end{minipage}
    
     \vspace{-10pt}
\end{figure*}

\begin{figure*}
\centerline{\includegraphics*[width=1\textwidth]{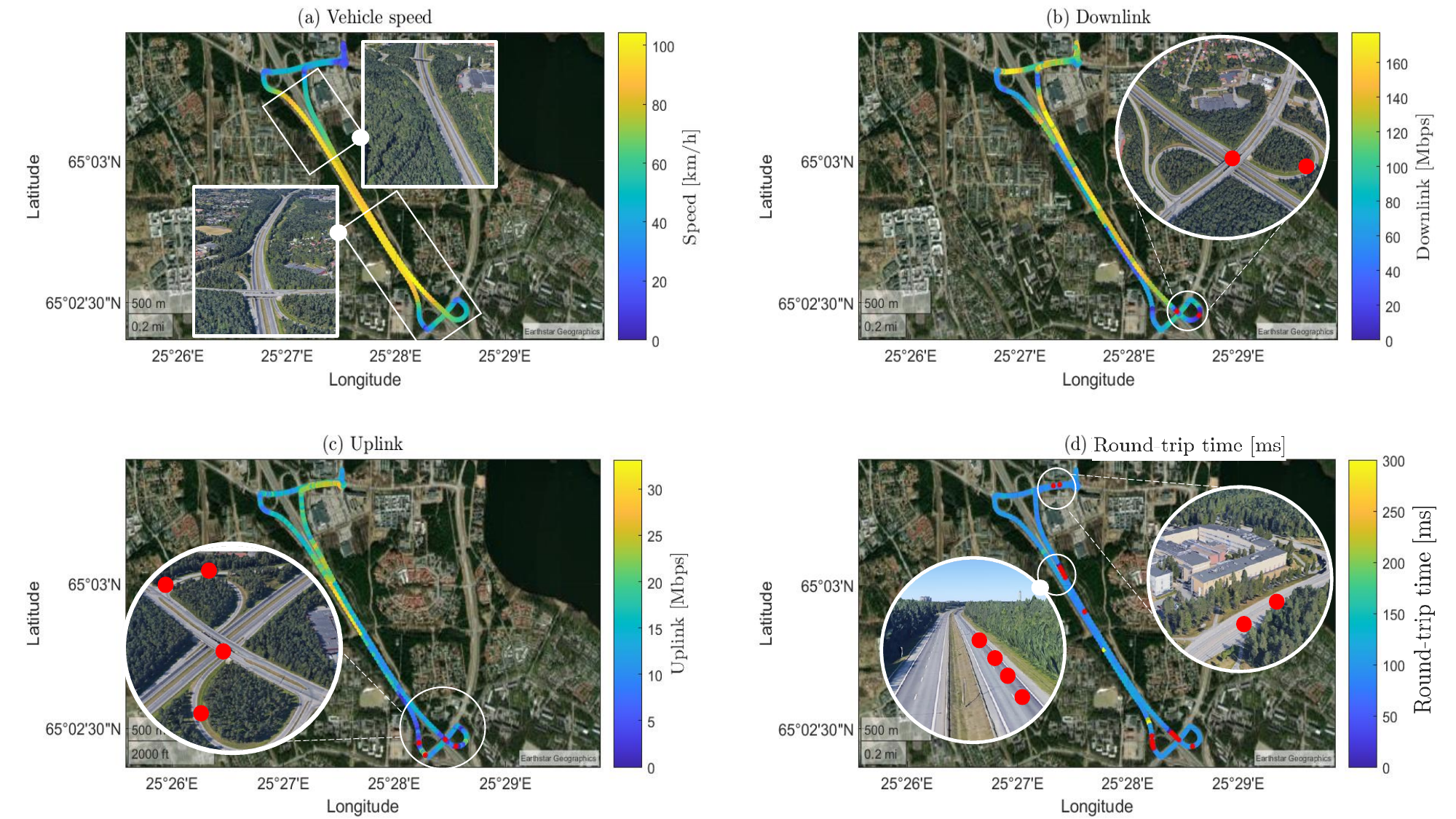}}
\caption{Starlink in-motion: (a) vehicle speed;  (b) downlink throughput; (c) uplink throughput and  (d) RTT, where outages are marked with red dots.} 
\label{fig:fig9}

 \vspace{-10pt}
\end{figure*}
\subsubsection{Stationary} Fig. \ref{fig:fig7} and Fig. \ref{fig:fig8} illustrate the Starlink uplink and downlink throughput performance. Specifically, Fig. \ref{fig:fig7} shows the Starlink downlink and uplink median throughput from four different test locations in Oulu to an \verb|iperf3| server in Finland. In the case of OuluZone, we use servers in Oulu and Helsinki. In Fig. \ref{fig:fig7} remaining measurements use only an \verb|iperf3| server in Helsinki. One can see that the measurement results are within the range of the Mobile Priority service plan. All test locations experienced median downlink and uplink throughput higher than 116 Mbps and 19.9 Mbps, respectively. The downlink performance for the Botanical Garden test location is significantly lower than the other test location due to possible LOS obstruction as indicated in Fig. \ref{fig:fig6}. However, this trend could possibly be due to the terminal being connected to a Starlink satellite in an unobstructed area of FoV.  The measurements at Auran Maja demonstrate the maximum downlink and uplink throughput 249 Mbps and 44 Mbps, respectively. Despite this, the median throughput remains closer to other other locations. Even though Mobile Priority data was already consumed,  the measurement results in Fig. \ref{fig:fig8} align closely with the Mobile Priority service plan. This indicates the absence of interference and network congestion by other Starlink users during the measurements in Oulu. Moreover, these results reveal that Starlink connection is the bottleneck, and server location does not affect the throughput.
\if{0}
\begin{figure}[t!]
\centerline{\includegraphics*[width=0.5\textwidth]{Median_All_Servers_B.eps}}
\caption{Starlink Downlink and uplink Internet speed from VTT's parking, Oulu to (i) Oulu (ii) Helsinki (iii) Amsterdam (iv) Frankfurt and (v) Reykjavík.}
\label{fig:fig8}
\end{figure}
\fi


\subsubsection{In-motion}
Fig. \ref{fig:fig10} compares the performance for stationary and in-mobile conditions. Fig. \ref{fig:fig9} dives deeper into the impact of vehicle speed on Starlink's RTT, downlink, and uplink throughput. In Fig. \ref{fig:fig10} the results indicate that the median of the downlink and uplink throughput decreased by 9 Mbps and 3.2 Mbps, respectively,  additionally, the median RTT increased by 9 ms when in motion. One of the primary reasons for this performance degradation is the roadside obstructions, e.g.,  trees and bridges. However, a larger dataset and more comprehensive analysis are needed to determine if other factors also contribute to the drop in performance while \textit{in motion}.
\if{0}
\begin{figure}[t!]
\centerline{\includegraphics[width=0.5\textwidth]{StationaryInMotion_D.eps}}
\caption{Comparison of Starlink downlink and uplink throughput and RTT for stationary and in-motion scenarios.}
\label{fig:fig10}
\end{figure}
\fi

Specifically, Fig. \ref{fig:fig9} (a) shows the vehicle speed for a single drive test between the European motorway E8 Exit 12 and Exit 11. One can see that on E8 the vehicle speed was stable around 100 km/h. However, the speed was under 60 km/h on the remaining path. Notably, the drive test path was mostly surrounded by trees as shown in Fig.~\ref{fig:fig9}~(a). For the three independent drive tests,  Fig.~\ref{fig:fig9}~(b) shows downlink, Fig.~\ref{fig:fig9}~(c) illustrates uplink, and  Fig.~\ref{fig:fig9}~(d) demonstrates RTT. These results demonstrate that Starlink is capable of supporting in-motion Internet connectivity, provided that the service can tolerate occasional outages. Since, 
a few connection outages were observed when the vehicle entered and left the motorway. Specifically, the curved on-ramps and off-ramps which were surrounded by dense trees caused LOS obstruction. Additionally, connectivity was lost when the vehicle passed beneath the bridge. These outages are marked with red dots in Fig. \ref{fig:fig9}. 

\section{Conclusion}
\label{sec:sec6}

This paper conducts measurements and discusses Starlink's performance for stationary and in-motion use in Northern Europe. Our results show that Starlink is capable of supporting both stationary and in-motion connectivity needs in Northern Europe. However, the performance fluctuates because of dynamic channel conditions and obstructions. We made our measurement dataset publicly available, which will facilitate the reproduction of results and support follow-up studies to conduct comparative analysis. Starlink terminals are designed to operate in harsh weather conditions, including snow, rain, and wind. In Finland, winter temperatures can drop as low as -$35^{\circ}\mathrm{C}$ to -$45^{\circ}\mathrm{C}$. Therefore, we consider it worth investigating how the different weather conditions such as summer rains and heavy snowfalls affect Starlink's performance. It would be good to study how real network payloads impact Starlink's latency, providing insights into its capabilities under real-world usage conditions. For future works,  it will be useful and interesting to expand these measurements to include multiple hardware terminals from different satellite internet providers and diverse environmental conditions to obtain a more comprehensive analysis. The research community has shown a growing interest in leveraging machine learning (ML) models for throughput prediction and data-driven optimization in satellite communications. We also recommend collecting a larger dataset for a more in-depth analysis and understanding of the Starlink routing strategy, as well as to support ML-based throughput prediction.     As another potential research direction, we propose examining the performance of a three-dimensional network that integrates terrestrial cellular networks and drones equipped with a Starlink mini terminal.

\section*{Acknowledgment}
This research was supported by Business Finland through the Drolo2 and 6G-SatMTC projects.
\vspace{12pt}
\bibliographystyle{IEEEtran}
\bibliography{IEEEabrv,ref}
\end{document}